# Twist-engineering of a robust Quantum Spin Hall phase in β-/flat bismuthene bilayer from first principles

Umberto Pelliccia,[a] Alberto M. Ruiz,[a] Diego López-Alcalá,[a] Gonzalo Abellán,[a] Rafael Gonzalez-Hernandez[b] and José J. Baldoví*[a]

[a] Instituto de Ciencia Molecular, Universitat de Valencia, Catedrático José Beltrán 2, 46980 Paterna, Valencia, Spain.

[b] Departamento de Física y Geociencias, Universidad del Norte, Barranquilla, Colombia.

E-mail: j.jaime.baldovi@uv.es

**Abstract:** Twist-engineering of topological phases in two-dimensional materials offers a powerful route to modulate electronic structure beyond conventional strain or chemical control. In particular, group 15 (pnictogens) monolayers such as bismuthene provide an ideal platform due to their strong intrinsic spin–orbit coupling (SOC) and robust topological character. Here, we investigate a previously unexplored heterostructure consisting of a β-bismuthene monolayer rotated by 30° on a planar bismuthene layer stabilized on a SiC(0001) substrate. Using first-principles calculations, we demonstrate that this specific rotational alignment induces a unique interlayer orbital hybridization which, combined with the strong SOC and the naturally broken inversion symmetry, gives rise to a pronounced Rashba spin-splitting, absent in the isolated monolayers. The topological nature of the system is confirmed through the calculation of the $Z_2$ topological invariant and Spin Hall Conductivity (SHC), revealing a robust Quantum Spin Hall (QSH) phase with an enhanced topological response compared to the individual layers. Furthermore, we explore the chemical tunability of this system via Sb substitution, showing that the gradual reduction of SOC systematically narrows the band gap while preserving the non-trivial topology. Our results establish large-angle twisted group 15 heterostructures as a versatile platform for engineering spin-orbit-driven phenomena and advancing topological spintronics.

## 1. Introduction

Since the discovery of graphene in 2004,[1] two-dimensional (2D) materials have been extensively explored for their broad range of electronic, optical and magnetic properties and their applications.[2–4] Beyond graphene-based materials, group 15 elemental monolayers have emerged as particularly promising due to the interplay between their 2D nature and relativistic effects.[5–10] In this context, bismuthene stands out as a prototypical platform due to its strong intrinsic spin–orbit coupling (SOC), which—unlike in graphene—opens a sizable topological energy gap governed by relativistic effects, enabling the realization of quantum spin Hall (QSH) phases.[11–15] This behavior has been explored in both bulk and reduced dimensionality systems. In three-dimensional compounds such as $Bi_2Se_3$ and $Bi_2Te_3$,[16,17] topological order manifests through gapless Dirac-like surface states protected by time-reversal symmetry and strong spin-momentum locking.[18–20] Furthermore, when structural inversion symmetry is explicitly broken, such as at interfaces or within heterostructures, these systems can develop Rashba spin-splitting.[21–23] In the 2D limit, bismuthene has been predicted and experimentally confirmed to host a QSH phase, particularly when stabilized as a planar monolayer on insulating substrates such as SiC(0001),[24] emerging as a robust 2D topological insulator.[13–15,24,25] Indeed, this flat configuration maximizes in-plane orbital hybridization, providing a highly tunable platform for controlling out-of-plane electronic coupling and breaking structural inversion symmetry,[24,25] thus making the system appealing for designing novel vertical heterostructures with tailored spin-orbital interactions.[26]

Beyond intrinsic material properties, recent advances have demonstrated that external degrees of freedom –such as strain, substrate interaction and stacking– can be exploited to further tailor electronic structures.[27–31] Among these approaches, twistronics has emerged as a powerful approach, where the relative rotation between adjacent layers introduces new periodicities and modifies interlayer coupling.[32–35] While most studies have focused on small-angle moiré superlattices, large-angle rotations can generate qualitatively distinct stacking configurations, leading to strong modifications of orbital hybridization and symmetry breaking.[36–38] In Bi-based systems, twist-induced phenomena remain comparatively unexplored, particularly in the large-angle regime. Previous works have shown that the impact of moiré patterns can modulate topological edge states and spin textures.[39] Furthermore, theoretical models of twisted bilayer bismuthene proved that the moiré twist acts as a tuning parameter to modulate SOC and shape the in-plane spin textures.[40–42] However, the role of commensurate high-angle configurations in driving new topological responses is still largely unknown. Moreover, combining structural engineering with chemical substitution offers an additional degree of control. In particular, alloying with lighter elements such as Sb provides a direct route to tune the strength of SOC and thereby the magnitude of the topological gap.[43,44]

In this work, we investigate a hybrid system composed of a zigzag β-bismuthene monolayer stacked on a planar bismuthene/SiC(0001) substrate with a relative twist angle of 30°. Using density functional theory (DFT), we analyze its structural stability, electronic properties and topological character. We demonstrate that the interplay between interlayer hybridization, strong SOC and broken inversion symmetry stabilizes a robust QSH phase with enhanced response. Finally, we show that this phase can be systematically tuned through Sb substitution, providing a flexible platform for engineering topological functionality in layered group 15 systems.

## 2. Computational methods

For the structural relaxation and the subsequent electronic structure calculations, we employed first-principles Density Functional Theory (DFT) as implemented in the Vienna ab Initio Simulation Package (VASP),[45] using the generalized gradient approximation (GGA). All the calculations were performed by including spin-orbit coupling (SOC). We employed projector-augmented wave (PAW) potentials[46] and the exchange-correlation potential is approximated with the Perdew-Burke-Ernzerhof (PBE) functional.[47] The plane-wave cutoff energy was set to 500 eV. We used the CellMatch code[48] to evaluate the lattice strain and the number of atoms for the commensurate supercell configurations. We sampled the Brillouin zone using an 11 x 11 x 1 Monkhorst-Pack $k$-point mesh. For the molecular dynamics thermal stability simulations, we instead used a 7 x 7 x 1 grid. The crystal structure was optimized until the forces on the ions were below 0.01 eV/Å and the total energy was converged to $10^{-5}$ eV with the Gaussian smearing method. To represent the phonon spectra, we used the Phonopy package[49] with a 3 x 3 x 1 supercell. We evaluated the thermal stability by performing ab initio molecular dynamics simulations (AIMD) within the canonical NVT ensemble. The simulations were carried out using the Nosé-Hoover thermostat[50,51] for a total of 5 ps, with a time step of 1 fs. We generated the inputs for these calculations using a tight-binding Hamiltonian obtained with the Wannier90 code,[52] employing a reduced basis set composed of the $p$ and $s$ orbitals of Bi, C and Si as well as the $p$ and $s$ orbitals of Sb in the doped systems. The WannierTools code[53] was employed to calculate the spin Hall conductivity (SHC) and topological $Z_2$ invariant. Computations were performed using a dense k-point grid of 121 × 121 × 1 with an energy interval of 1.0 eV. Furthermore, the out-of-plane lattice parameter $c$ was set to 30 Å to prevent spurious interaction between periodic images.

## 3. Results and discussion

First, we investigate the structural and electronic properties of the individual building blocks that constitute the twisted β-/flat bismuthene bilayer. The planar bismuthene monolayer on a SiC(0001), hereafter denoted as Bi/SiC, adopts a fully planar geometry, commensurate with the substrate in a $(\sqrt{3} \times \sqrt{3})$R30° configuration, in agreement with previous reports.[24] Our structural optimization yields lattice parameters of $a = b = $ 5.394 Å, corresponding to an in-plane Bi-Bi distance of 3.090 Å. In contrast, free-standing zigzag β-bismuthene monolayer stabilizes in a buckled honeycomb structure, characterized by lattice parameters of $a = b = $ 4.336 Å, a Bi-Bi distance of 3.047 Å and a vertical buckling height of 1.746 Å. This structural distinction reflects the intrinsic competition between $sp^2$-like planar bonding and $sp^3$-like hybridization in group 15 elemental monolayers, which becomes a key factor when interfacing both phases.

To construct the heterostructure, we stack the zigzag β-bismuthene monolayer on top of the planar Bi/SiC substrate with a relative twist (rotation) of 30° (Fig. 1(a-b)). This specific angle enables a commensurate supercell with minimal lattice mismatch, requiring a moderate tensile strain of ~6% on the zigzag Bi layer (Tab. S1) and resulting in lattice parameters of $a = b = $ 9.264 Å. The strain is predominantly accommodated by the more flexible zigzag layer, while the Bi/SiC is comparatively more rigid and is not expected to undergo significant deformation upon deposition of an isolated monolayer.

The dynamical stability of the resulting structure is confirmed by phonon calculations (Fig. 1(c)), which show the absence of imaginary frequencies across the Brillouin zone. Furthermore, the vibrational modes reveal a mass-dependent separation: low-energy modes are purely Bi-derived, intermediate frequencies correspond to Si and C substrate vibrations while the frequencies above the gap are mainly governed by H with some contribution from C. In addition, ab initio molecular dynamics (AIMD) simulations demonstrate thermal stability up to 300 K, with no structural degradation observed, whereas partial disorder emerges at 600 K (Fig. 1(d)). These results establish the robustness of the heterostructure under ambient conditions.

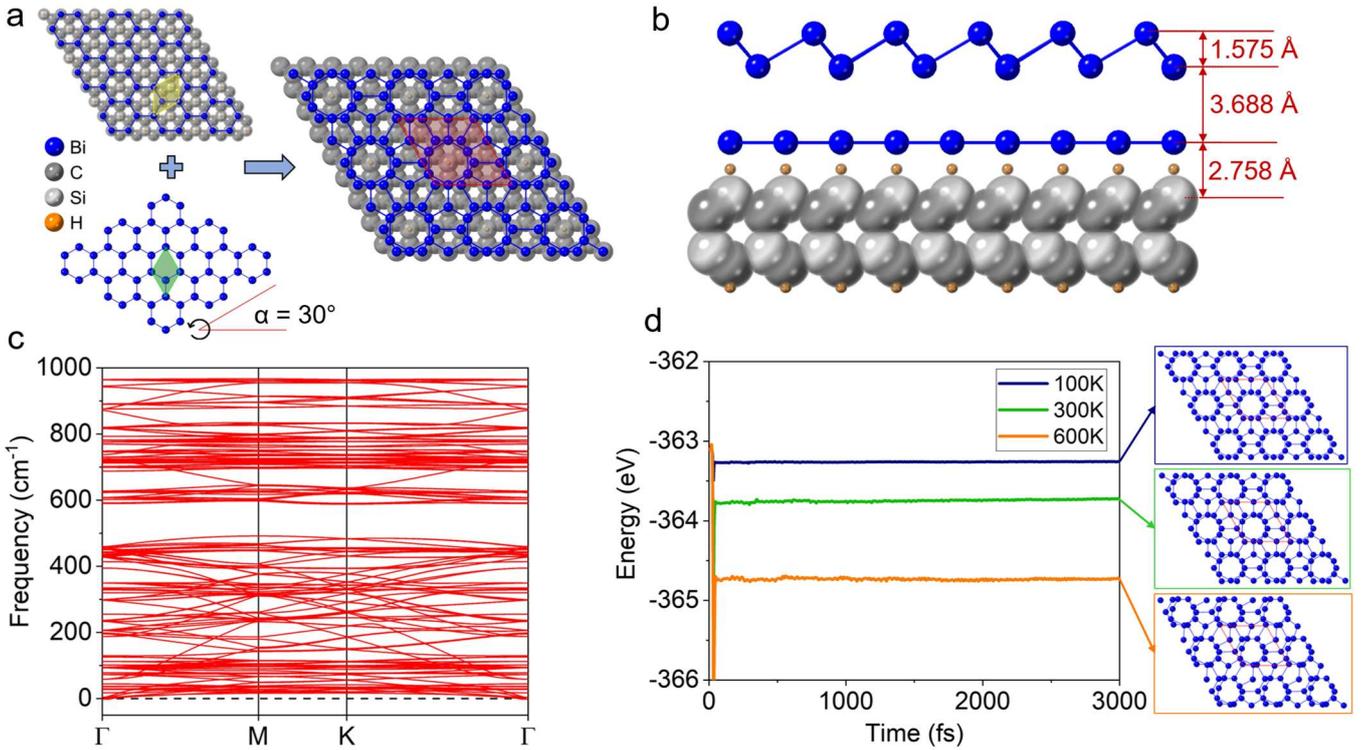

**Fig. 1** (a) Top and (b) lateral view of atomic structure models of the moiré heterostructure formed by a 30° twisted zigzag β-bismuthene monolayer stacked over Bi/SiC. (c) Phonon dispersion curve of the heterostructure. (d) Ab-initio molecular dynamics simulation snapshot at 100K, 300K and 600K.

We now turn to the electronic properties. The pristine bismuthene monolayer, in its freestanding form, exhibits a band gap of 0.452 eV at the $\Gamma$ point (Fig. 2(a)). However, when integrated into the commensurate supercell, this gap is reduced to 8.9 meV at the $\Gamma$ (Fig. S1(a)) in line with previous theoretical predictions.[54] Despite this reduction, the imposed strain is physically justified by the interlayer interaction within the heterostructure. Indeed, the optimized interlayer distance of 3.688 Å (Fig. 1(b)) is notably smaller than the sum of the van der Waals radii of Bi atoms (~4.14 Å). This reduced separation indicates the emergence of partially covalent, or metavalent[55,56] bonding between the two layers. This promotes a structural rearrangement of the zigzag β-bismuthene to match the Bi/SiC lattice. To further ensure the robustness of our results, we allow the heterostructure to relax starting from the optimized lattice constant of free-standing zigzag β-bismuthene ($a = b = 8.7$ Å), which effectively imposes compressive strain on the Bi/SiC substrate. Under these conditions, the system evolves into a configuration in which Bi/SiC remains essentially undistorted with $a = b = 9.305$ Å and in turn imposing tensile strain to the β-bismuthene monolayer due to its higher flexibility (Tab. S2).

Next, we evaluate the electronic properties of the heterostructure. Before integration, the planar Bi/SiC system exhibits a topological insulating phase[24] with an indirect band gap of 0.488 eV (Fig. 2(b)), where the valence band maximum (VBM) is located at the K point, while the conduction band minimum (CBM) resides at $\Gamma$. This is in qualitative agreement with the reported theoretical band gap of 0.67 eV.[24] Upon formation of the $(\sqrt{3} \times \sqrt{3})$R30° supercell, the resulting Brillouin-zone folding maps the VBM from K to $\Gamma$ (Fig. S2(b)), enabling a direct energetic overlap between the valence states of the planar layer and the conduction states of the zigzag layer. As a consequence, the twisted heterostructure develops a direct band gap of 50.6 meV at the $\Gamma$ point (Fig. 2(c)), representing a non-trivial reconstruction of the electronic structure driven by interlayer coupling.

In order to rationalize the effect of SOC in the band structure, we compare the electronic band structure both with and without SOC. In the absence of SOC, the system exhibits a metallic character with a Dirac-like crossing point at $\Gamma$ (Fig. S2(c)). The inclusion of SOC lifts this degeneracy and opens a gap, indicating that the insulating state is entirely driven by relativistic effects. In contrast to graphene-like systems, where SOC acts as a weak perturbation, here the gap originates from strong on-site SOC acting on Bi p orbitals, leading to a hybridization gap. The observed opening of the hybridization gap at the $\Gamma$ point induced by strong SOC suggests that the system hosts a non-trivial topological phase. To characterize this phase, we computed the $Z_2$ topological invariant by means of the Wannier charge center analysis.[53] Our calculations confirm that the twisted bismuthene heterostructure preserves the topological insulating state inherited from the free-standing zigzag β-bismuthene layer, with a nontrivial $Z_2$ invariant ($v = 1$) consistent with previous calculations on ultrathin Bi(111) films.[15] This topological character is further reflected in the finite spin Hall conductivity (SHC) exhibited by the β-/flat bismuthene heterostructure and its constituent layers, as shown in Fig. 2. Specifically, the SHC values within the band gap are 1.75 (e/4π) for the β-/flat bismuthene heterostructure, compared to 1.12 and 1.03 for the Bi/SiC and bismuthene monolayer, respectively, which indicates an enhanced topological response arising from interlayer hybridization.

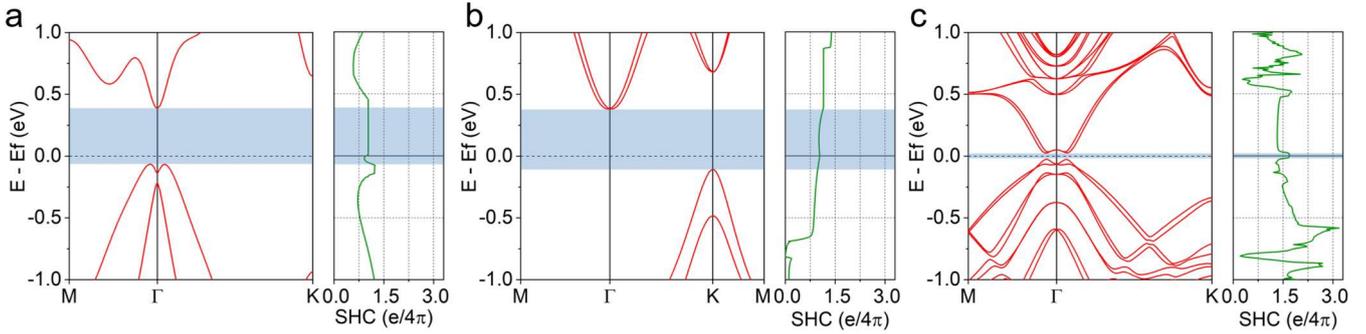

**Fig. 2** Calculated electronic band structures and corresponding SHC spectra for (a) the free-standing bismuthene monolayer, (b) the planar bismuthene over SiC substrate (Bi/SiC) and (c) the twisted heterostructure. The blue shaded regions highlight the energy window of the band gaps. Interlayer hybridization and strong SOC in the heterostructure induce a direct band gap of 50.6 meV at Γ point, alongside an enhanced topological response.

A detailed analysis of the layer-resolved projected band structure (Fig. 3(a–d)) reveals that the electronic states near the Fermi level (±0.1 eV) originate from a strong hybridization between the two Bi layers. These states are predominantly composed of Bi $p$ orbitals (Fig. S4), whereas the SiC substrate contributes negligibly in this energy window, confirming that the low-energy physics is governed by the bilayer system. Furthermore, in addition to the topological gap, the broken inversion symmetry of the heterostructure gives rise to a pronounced Rashba spin splitting near the VBM. The spin-resolved Fermi surface (Fig. 3(e)) displays a characteristic helical texture, with in-plane spin components locked perpendicular to momentum. This behavior reflects the combined effect of strong SOC and structural asymmetry and provides further evidence of spin-momentum locking in the system.

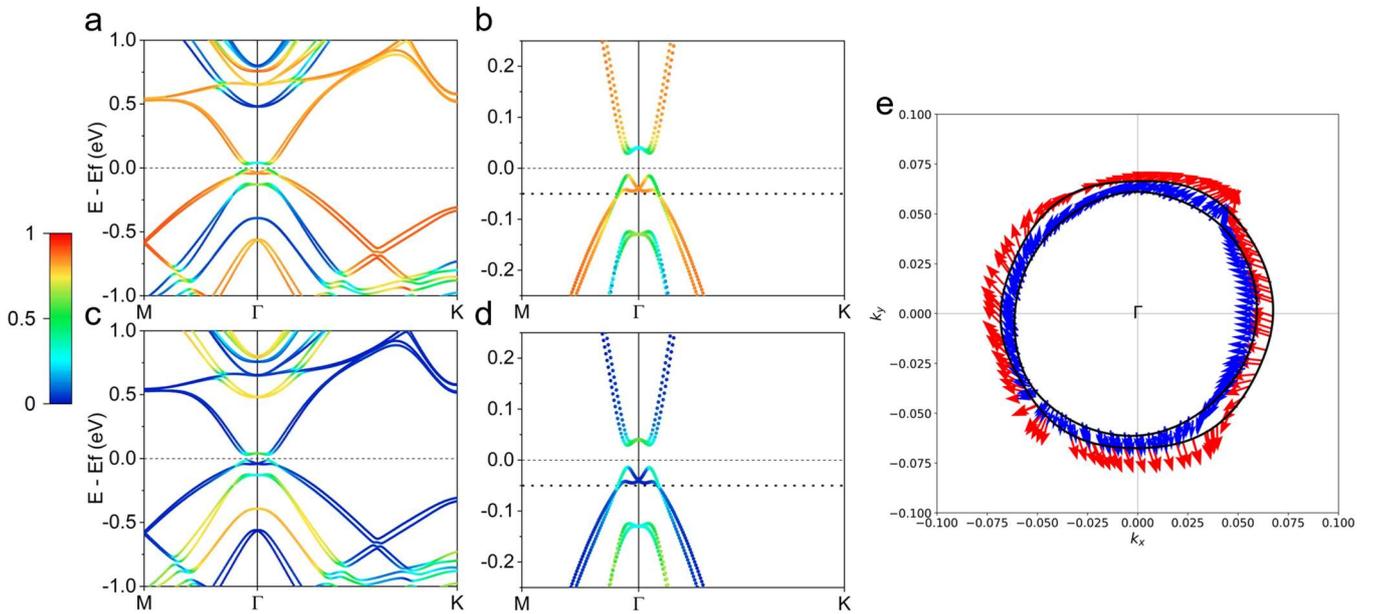

**Fig. 3** Layer-resolved projected band structure of the twisted heterostructure for (a) the zigzag β-bismuthene monolayer and (c) the planar bismuthene layer. (b, d) Corresponding zoomed-in views near the Fermi level. The color gradient represents the fractional projection of the electronic states onto the atomic orbitals of the respective layer, ranging from red (high contribution, 1.0) to blue (no contribution, 0.0). (e) Band-resolved spin texture calculated at -0.05 eV relative to the middle of the band gap. The arrows illustrate the in-plane spin direction, with red and blue colors distinguishing the two split bands to highlight the spin-momentum locking.

Finally, we explore the chemical tunability of the heterostructure through substitutional Sb doping in the zigzag layer, which shares the same valence configuration ($s^2p^3$) as Bi but has a smaller atomic radius and weaker SOC. The experimental feasibility of these systems can be envisioned by taking advantage of ultra-high-vacuum methodologies (e.g., epitaxial growth, chemical vapor transport, or pulsed laser ablation), as well as recent developments in bottom-up wet-chemical synthesis, which have provided versatile and scalable alternatives that enhance their technological viability.[10,57–59] Therefore, we construct four distinct doped configurations corresponding to Sb concentration of 25%, 50%, 75% and a fully substituted 100% phase within the zigzag layer, i.e. $Bi_{1-x}Sb_x$ on Bi/SiC where $x$ = 0.25, 0.50, 0.75, 1.00, respectively. To properly model the doped heterostructures, we evaluate the total energies of various arrangements. For each Sb concentration, we systematically select the ground-state configuration with the lowest energy (Tab. S3). Notably, the most stable structures are the ones that have the Sb atoms replacement in the bottom sub-lattice, which is directly adjacent to the flat bismuthene interface. Fig. 4 illustrates the corresponding optimized geometries selected for the different doping levels. The incorporation of Sb atoms leads to a progressive increase of the vertical distance between the doped zigzag layer and the flat bismuthene, going from 3.698 Å ($x$ = 0) to a maximum 3.764 Å ($x$ = 1.00). This expansion occurs because the smaller Sb orbitals weaken the attractive interlayer hybridization, allowing the layers to slightly decouple.

The electronic band structure of the doped systems (Fig. 4(a-d)) shows that the direct gap at $\Gamma$ is preserved across all concentrations (Fig. S8), although its magnitude decreases monotonically from 50.6 meV to 16.8 meV as $x$ approaches 1.00. This trend directly follows the reduction in SOC strength associated with Sb substitution. Importantly, the orbital character near the Fermi level remains dominated by hybridized $p$ orbitals of Bi and Sb depending on their concentration in the system, indicating that the fundamental mechanism governing the electronic structure is retained. Contributions from the SiC substrate remains negligible, confirming that these tunable electronic features are strictly confined to the bilayer heterostructure (Fig. S5).

Despite the reduction in gap size, the topological phase remains robust throughout the entire Sb doping range, as confirmed by both the $Z_2$ invariant and the SHC calculations (Fig. 4). Specifically, the SHC increases from 1.83 $e/4\pi$ at 25% Sb doping to a maximum value of 2.1 $e/4\pi$ in the fully substituted phase ($x$ = 1.0), with intermediate values of 1.6 $e/4\pi$ at $x$ = 0.5 and 1.9 $e/4\pi$ at $x$ = 0.75. The observed enhancement in SHC, despite the reduction in band gap magnitude, can be attributed to subtle changes in the band topology induced by alloying, which modify the distribution of Berry curvature in momentum space. Overall, these findings highlight that chemical substitution with Sb serves as an effective and tunable parameter to optimize SOC within the twisted bismuthene bilayer heterostructure.

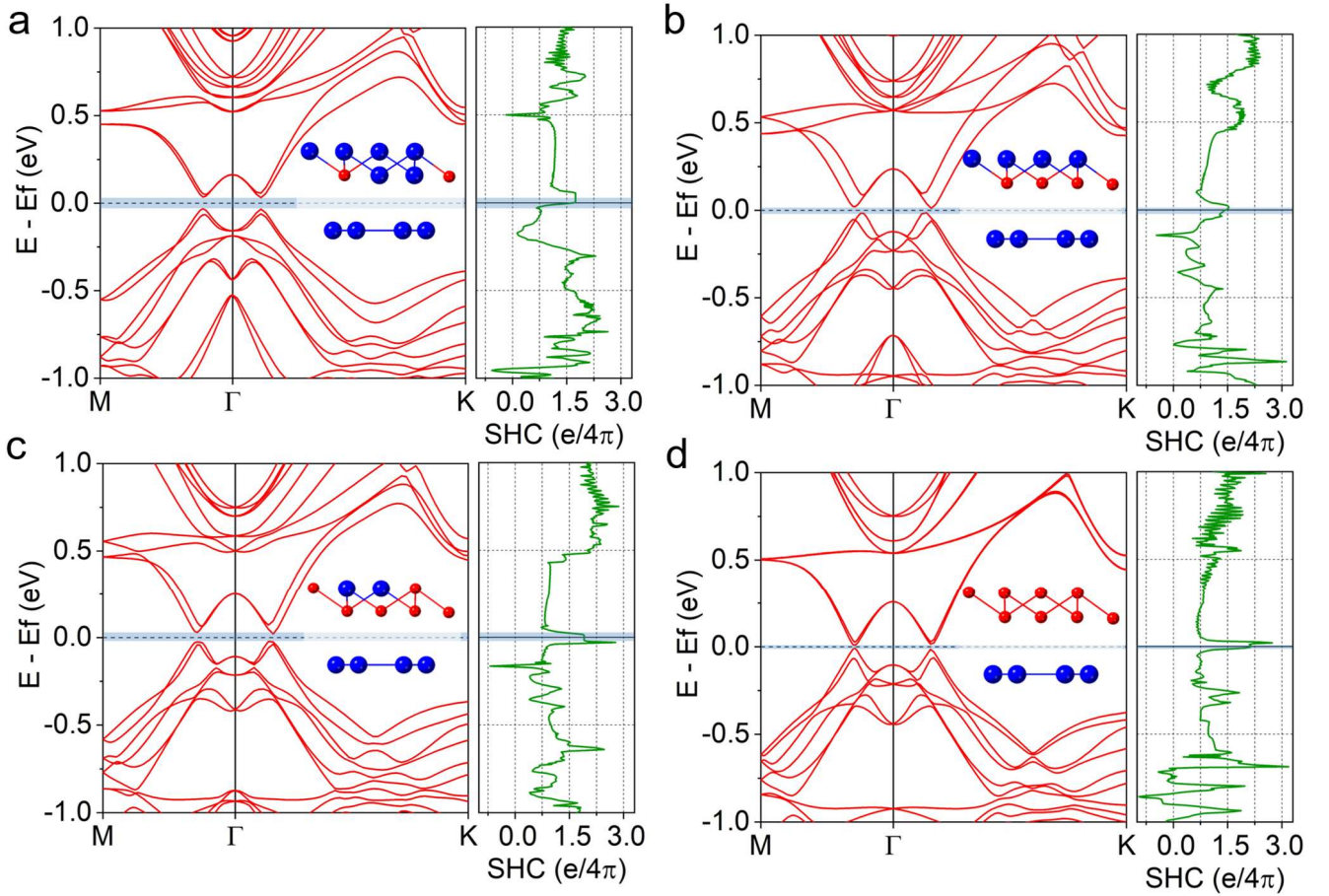

**Fig. 4** Calculated electronic band structures and corresponding SHC spectra for the Sb-doped heterostructure $Bi_{1-x}Sb_x$ on Bi/SiC at varying concentrations: (a) $x$ = 0.25, (b) $x$ = 0.50, (c) $x$ = 0.75 and (d) $x$ = 1.00. The insets display the relaxed atomic configurations for each doping level. The blue shaded regions highlight the band gap, which systematically decreases in magnitude. Furthermore, The SHC spectra confirms the persistence of the non-trivial topological phase, with the response peaking at $x$ = 1.00.

## 4. Conclusions

In summary, we have demonstrated that large-angle twist engineering provides an effective route to stabilize and enhance topological phases in group 15 heterostructures. By constructing a 30° twisted bilayer composed of a zigzag β-bismuthene monolayer on a planar Bi/SiC(0001) substrate, our first principles calculations reveal that this specific rotational alignment induces a unique interlayer orbital hybridization. This, combined with the strong SOC and the naturally broken spatial inversion symmetry of the heterostructure, drives the emergence of Rashba spin-splitting in the electronic bands. This feature is notably absent in the isolated monolayers. Furthermore, calculation of the $Z_2$ topological invariant and the SHC confirm that the twisted system hosts a robust QSH phase with an enhanced topological response relative to the individual constituent layers. Additionally, we systematically investigated the effect of chemical substitution with Sb at various concentrations. Our results demonstrate that substituting Bi with lighter atoms effectively modulates these interlayer interactions, leading to a systematic reduction of the band gap as the Sb concentration increases, without breaking the nontrivial topological state. These findings highlight this large-angle twisted heterostructure as a versatile platform in which topology, spin texture and electronic structure can be engineered for the development of tuneable spintronic devices.

## Author contributions

This work is part of the PhD thesis of U.P. U.P. performed the DFT calculations with the assistance of A.M.R. and D.L.A. U.P. and A.M.R. analyzed the data and interpreted the results. R.G.H. performed the SHC and topological $Z_2$ invariant calculations. G.A. supervised the chemical feasibility and analyzed the results. J.J.B. conceived and supervised the work. The manuscript was written by U.P., A.M.R., R.G.H. and J.J.B. All authors have given approval to the final version of the manuscript.

## Conflicts of interest

There are no conflicts of interest to declare.

## Acknowledgements

The authors acknowledge financial support from the European Union (ERC-2021-StG101042680 2D-SMARTiES), the Spanish Government MCIU (PID2024-162182NA-I00, PID2022-143297NB-I00), the Generalitat Valenciana (grants CIDEXG/2023/1, CIDEGENT/2018/001) and the Spanish MICINN (Excellence Unit "Maria de Maeztu" CEX2024-001467-M). A.M.R. thanks the Spanish MIU (grant no. FPU21/ 04195). The calculations were performed on the HAWK cluster of the 2D Smart Materials Lab hosted by Servei d'Informàtica of the Universitat de València.